\documentstyle[aps,prl,epsf,rotate]{revtex}   

\begin{document}

\title{Fast diffusion of a Lennard-Jones cluster on a crystalline surface}

\author{Pierre Deltour, Jean-Louis Barrat and Pablo Jensen}

\address{
D\'epartement de Physique des Mat\'eriaux, Universit\'e Claude Bernard
Lyon-1, CNRS UMR 5586, 69622 Villeurbanne C\'edex, France}

\maketitle

\begin{abstract}

We present a Molecular Dynamics study of  large Lennard-Jones clusters evolving on a 
crystalline surface. The static and the dynamic properties of the cluster are described.
We find that large clusters can diffuse rapidly, as experimentally observed.
The role of the mismatch between the lattice parameters of the cluster and 
the substrate
is emphasized to explain the diffusion of the cluster.
This  diffusion can be described as a Brownian motion
induced by the vibrationnal coupling to the substrate,
a mechanism that has not been previously considered for cluster diffusion.

\end{abstract}


Understanding the interaction of particles of nanometer size with surfaces
is important, both from a fundamental point of view and for controlled growth
of thin films and nanostructures.  Recently,
the diffusion of large clusters containing hundreds of atoms was investigated
 both
experimentally \cite{diffexp} and theoretically \cite{diffth}. These studies
have focused on clusters epitaxially oriented on the surface, and have found
relatively low diffusion coefficients, of order $10^{-17} cm^2 s^{-1}$.
In contrast, Bardotti et al. \cite{prl} have shown experimentally 
that large (one hundred to a few thousand atoms) gold or antimony clusters, 
non-epitaxially oriented, have a surprisingly high diffusion coefficient 
(of the order $10^{-8} cm^2 s^{-1}$ at room temperature) on a graphite substrate.  
This observation is reminiscent of early work by Kern et al. \cite{Kern}, in which 
a noticeable diffusion phase preceding epitaxial locking was observed for
small gold crystallites on a $NaCl$ surface.  The essential conclusion that 
can be drawn is that "large" (a few nanometers) objects 
can have large surface diffusion coefficients at room temperature,  
a result that might seem rather counterintuitive  to many surface scientists.
Indeed, most diffusion mechanisms that have been 
considered for surface diffusion of composite objects such 
as clusters involve a combination  of single atom diffusion events
(e.g.  evaporation condensation \cite{diffth}), and yield diffusion 
constants much smaller than the above mentioned result. Although
these mechanisms have been shown to describe correctly the diffusion
of clusters of adatoms in epitaxy on the substrate, they do not seem 
to be relevant to explain the fast diffusion that is observed in references
\cite{prl,Kern}. 

In this Letter, we present the results of a molecular
 dynamics (MD) study of the cluster diffusion problem, with particular 
emphasis on the consequences of epitaxial or non-epitaxial cluster-substrate
configurations. The non-epitaxial
case is important for growth of non-epitaxial thin films or 
for films prepared by {\it pre-formed} cluster deposition \cite{review}.
 A  simple model, aimed at clarifying the generic aspects of the question rather
than modeling a particular case, was used.  Both the cluster and the 
substrate are made up of Lennard-Jones atoms,  interacting through 
potentials of the form
$
V(r)=4 \epsilon \left (\left(\frac{\sigma}{r} \right)^{12} 
- \left(\frac{\sigma}{r} \right)^{6} \right )
$. 
Empirical potentials of this type, originally developed
for the description of inert gases,  are now commonly
used to model generic properties of condensed systems.
The substrate is modeled by a single layer 
of atoms on  a triangular lattice, attached  to their 
equilibrium sites  by weak harmonic springs that preserve surface
cohesion.  The Lennard-Jones parameters for 
 cluster atoms and substrate atoms are
respectively $\left (\epsilon_{cc},\sigma_{cc} \right) \mbox{ and }
\left (\epsilon_{ss},\sigma_{ss} \right)$. The parameters
$\left (\epsilon_{sc}, \sigma_{sc} \right)$ for the substrate-cluster
 interaction are
constructed by following the standard combination rule : $ \epsilon_{ss} 
 \sim  \sigma_{ss}^{6} $ and 
$ \sigma_{sc} =  \frac {1}{2} \left ( \sigma_{cc} + \sigma_{ss} \right )$.
 Finally,  the unit of time is defined as 
 $\tau = (M \sigma_{cc}^2/\epsilon_{cc})^{1/2}$, where $M$ is the mass of the 
atoms which is identical for cluster and substrate atoms.
 
Our simulation uses a standard   molecular dynamics technique
with thermostatting of the {\it surface} temperature
 \cite{A&T}. 
The equations of motion are then integrated 
using the Leap-Frog Algorithm\cite{A&T}, which is a simple, time
reversible and very stable integration scheme \cite{Frenkel}
The run  is started  from a configuration in which
the cluster atoms occupy  sites  of a cubic lattice near
the substrate. The system is then equilibrated for 500$\tau$,
after which the trajectories are recorded. Diffusion constants $D$ 
are obtained from the mean-squared displacement 
of the cluster center of mass.
Typically, the length of the runs used to compute $D$ was 
 2500 $\tau$, with some longer runs of  12500 $\tau$. This means 
that diffusion constants smaller than $5\ 10^{-4}\sigma_{cc}^2\tau^{-1}$,
for which the cluster does not visit an area of more than $25\sigma_{cc}^2$
 over the length of the simulation,  are not accessible.

We first briefly describe the static properties of
our supported clusters. At the temperature at which most runs were
carried out, $T=0.3\epsilon_{cc}$,  the clusters are in  a crystalline
FCC or HCP configuration.  They take
the spherical cap  shape of a solid droplet (figure 1) partially wetting the 
substrate. The contact angle $\theta$, that can be defined following reference
 \cite{angle}, is roughly independent of the 
cluster particle number  $N$, for $50 < N < 500$. This angle can be changed 
by tuning the cluster-substrate interaction. For large enough 
$\epsilon_{sc}$, total wetting is observed, and the cluster dissociates.
As noted in \cite{angle}, this shows the relevance of 
macroscopic concepts such as contact angle and wetting even for 
nanometer-sized particles.

We now turn to the dynamical behavior of the 
supported cluster.  Most of the runs were
carried out at a reduced temperature of 0.3 so that
the cluster is clearly  solid.  In that case, visual observation
of atomic trajectories during the run indicate that the 
diffusion inside the cluster, or on its surface, is completely negligible.
Except for the vibrational motion of the atoms, 
the cluster behaves as a rigid object.  This is clearly visible in figure 
1, where the left and right halves of the cluster,  colored  
grey and white  at 
the beginning of the run, clearly retain their identity after the 
cluster {\it center of mass}  has moved over 2 lattice parameters. Hence the 
  motion of the cluster appears to be 
 controlled by collective motions of the cluster as a whole rather than
by single atomic jumps.  This collective diffusion 
mechanism will 
 depend essentially on three parameters: the lattice parameter 
of the substrate, the temperature and the cluster size. We consider in
turn the influence of each parameter.

We first investigate the effect of the ratio of the cluster lattice parameter to the substrate lattice parameter 
for $T=0.3 \epsilon_{cc}$ and $N=100$,  by varying the interaction diameter of the substrate
atoms in the range of $0.5\sigma_{cc}$  to  $1.5\sigma_{cc}$
The lattice constant $d$ of the substrate atoms is also correspondingly  scaled in such
a way that $\sigma_{ss}/d$ remains constant, equal to its value
for a Lennard-Jones solid at zero pressure.  The results for
the diffusion coefficient are shown in figure 2.
When the substrate and cluster are commensurate ($\sigma_{ss}=\sigma_{cc}$),
 the cluster 
can lock into a low energy epitaxial configuration. A global translation
of the cluster would imply overcoming an energy barrier scaling
as $N^{2/3}$, the contact area between the cluster and the substrate.
 In that case,  single atom mechanisms as described in \cite{diffth} will
dominate, and the diffusion will be very slow. Indeed on the time scale of the 
MD simulations, diffusion was vanishingly small in that case. The points in
figure 2 at $\sigma_{ss} \sim \sigma_{cc}$ do not in fact represent 
 true diffusion constants, since in these cases the center of mass did not move
by more than a few lattice constants over the whole duration of the run.
However, for small deviations from this commensurate case,
the diffusion becomes measurable on the time scale 
of the MD runs.  This can be understood from the fact that the effective
potential in which the center of mass moves is much weaker, as the cluster 
atoms, constrained to their lattice sites inside the rigid solid cluster, are
 unable to adjust to the substrate potential.  The effect is rather spectacular.
A fifteen percent change on the lattice parameter induces a two order of
magnitude change on the diffusion coefficient. For the same reason, the diffusion
coefficient decreases slightly when $\sigma_{cc}$ is getting close to $2 \sigma_{ss}$.
Finally, we note that the effect is not exactly symmetric on both sides of the
commensurate situation. Clusters with atoms smaller than the substrate atoms
tend to diffuse more slowly, since the potential wells for these atoms will
be  deeper than for big adatoms.

Next, we study the influence of temperature for a fixed values
of $\sigma_{ss}$ and $\epsilon_{sc}$,  $\sigma_{ss}=0.9$ and $\epsilon_{sc}=0.4$. The diffusion constant 
can be fitted by an Arrhenius law (figure 3) , with  an  activation energy E of
 $0.66 \epsilon_{cc}$ and a prefactor 
$D_0$ of $0.02 D_\star$ where $D_\star$ is equal to $\sigma_{cc} \left( \epsilon_{cc} / M \right )^{1/2}$.
 However the
mechanisms involved here are not simple, single atom  activated processes. Not too much meaning should then be given to the 
activation energy and prefactor.

In order to establish a connection with experiments, we can identify 
$\epsilon_{cc}/k_B$
to a typical  melting temperature (1000K for Au).
Using  $\sigma_{ss} \sim 5$ \AA\  and $M\sim 10^{-25} kg$, we find that
our diffusion coefficients would fall in the range of
 $10^{-5} cm^2 s^{-1}$ for a 100 atoms cluster. This is in reasonable
agreement with the values obtained in \cite{prl}
, but differs strikingly
from the values obtained with single atom mechanisms. 

Finally, for the same values of $\epsilon_{sc}$ and $T$, 
 the effect of cluster size on the diffusion constant is considered for different
lattice parameter values.
As the number $N$ of atoms in the cluster is varied between
$N=10$ and $N=500$, the diffusion constant decreases, roughly following 
 a power law $D\sim N^{\alpha}$ (figure 4).
This power law exponent $\alpha$ depends significantly on the mismatch 
between the cluster and the substrate lattice parameters. 
For high mismatches ($\sigma_{ss}=0.7,0.8$), $\alpha$ is 
 close to $-0.66$.
 As the diffusion constant is
inversely proportional to the cluster-substrate friction coefficient, 
this result 
is in agreement with a simple "surface of contact"
argument  yielding  $D\sim N^{-2/3}$. On the other hand,
 when the lattice mismatch is equal to $0.9$, $\alpha$ one obtains 
$\alpha \approx -1.4$, 
although the shape of the cluster, characterized 
by the contact angle, does not appreciably change.
Moreover, the trajectory followed by the the cluster center of
mass changes qualitatively. In the runs with a large mismatch, this trajectory
is brownian like, with no apparent influence of the substrate. 
 In contrast, when the mismatch
is small, the center of mass of the 
cluster follows a "hopping-like" trajectory,  jumping 
from site to site on the honeycomb lattice 
defined by the substrate . When $\sigma_{ss}=\sqrt{3}/2$, there seems to be  a transition 
between the two regimes around $N=200$.

In an attempt to disentangle 
the contribution of the cluster internal vibrations to the diffusion process
from that
of the substrate vibrations in the small mismatch case ($\sigma_{ss}=0.9$),
 we now consider 
 idealized cases in which one of the two subsystems is 
artificially "frozen", so that its contribution vanishes.
First,  we consider the case of  cluster supported by a "frozen" 
substrate, with atoms constrained to their equilibrium 
positions. The external potential experienced
by the cluster atoms is thus purely static.  The cluster is first equilibrated 
using constant temperature MD at $T=0.3\epsilon_{cc}$, and this equilibration 
period 
is followed by a constant energy simulation.  The results for the diffusion
 constant are  very close to those obtained 
with a thermalized substrate. This shows that such clusters have enough 
internal degrees of freedom to play the role of being their own thermostat, 
and that their internal vibration modes can be an efficient motor
for the diffusion.  

Next, we consider the  other extreme case of a  "frozen" cluster deposited 
on a thermalized substrate. The cluster is first equilibrated,
either on a perfectly flat substrate exerting an average potential 
equivalent to that of the triangular lattice of  Lennard-Jones atoms, or 
in free space.  In the first case, it adopts the usual spherical cap shape, 
while in the
second case a quasi-spherical, faceted shape is observed.  This last shape 
would be reasonable for a cluster made of a highly cohesive material. After 
this equilibration phase, the cluster is "frozen" and deposited on the
 thermalized 
substrate as a  solid body. The center of mass trajectory is integrated   
using the quaternion algorithm \cite{A&T}.

 Not surprisingly, 
the diffusion is dependent on the way the system has been equilibrated.
For a spherical rigid cluster, the system rotates until a 
facet comes into contact with the substrate, then diffuses 
without rotating for the rest of the MD run.  Hence 
despite the more or less spherical shape of the cluster, rotation 
does not seem to give an important contribution to the 
diffusion.  In that case, 
the diffusion depends on the shape and size of the facet the cluster rests on,
 so that the results in this case are not well reproducible.  
For  a  cluster equilibrated on a flat surface, the diffusion constant  follows
the same power law as in the free cluster case.
 This surprising result suggests
 that the diffusion mechanism in our case cannot
be just simply explained in terms of dislocation migration within the cluster as
proposed to explain the diffusion of 2D islands in \cite{Hamilton}. Rather, the 
motor for diffusion is here the vibrational motion of 
the substrate, and its efficiency appears to be comparable to that 
of the internal cluster modes.

In summary, the surprisingly high diffusivity of large clusters can be
understood with two main ingredients.
When the cluster is not commensurate 
with the substrate, 
the modulation of the  potential felt by the center 
of mass is small. In this case the cluster is not locked by the
substrate and can vibrate relatively freely. The vibrations (phonons)
 of the substrate and the internal vibrations of the cluster both
create a "random" force on the cluster center of mass, which executes
a brownian motion in this weak external potential.  The two components 
of this random force appear to have similar intensities, and, at the temperature
we consider, are sufficiently strong to overcome the small energy barriers,
resulting in a rapid diffusive motion. Our calculation demonstrates
the efficiency of this brownian-like   mechanism, which was not 
considered in earlier studies of cluster diffusion, for explaining
the diffusion of rather large objects.
Further analysis of the vibrational
coupling between the substrate and the cluster will be necessary to
 fully understand the  mass dependence of the diffusion constant.

We thank the Centre Pour le D\'eveloppement du Calcul
 Scientifique
Parall\`ele at Universit\'e Claude Bernard Lyon-I
 for the allocation of computer 
time that made this work possible.

\begin{figure}
\vspace*{2cm}
\epsfxsize=6cm
\rotate[r]{\epsfbox{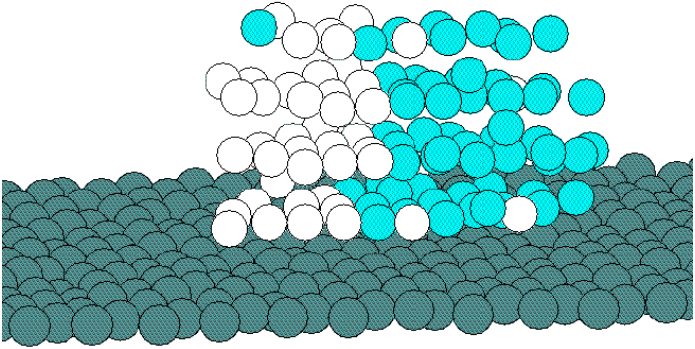}}
\caption{Configuration of the cluster on the crystalline surface. The cluster
is partially wetting the surface. 
The left and the right part of the cluster have been colored at the beginning of
the run. After the cluster center of mass has moved by two substrate
lattice constants from its original position,  the two parts of the cluster are
still well distinct. The cluster diffusion can then not been explained
in terms of single atom mechanisms.(N=100, $\sigma_{ss}$=0.9, $\epsilon_{sc}$=0.4, T=0.3)
\label{}
}
\end{figure}
\begin{figure}
\vspace*{2cm}
\epsfxsize=6cm
\rotate[r]{\epsfbox{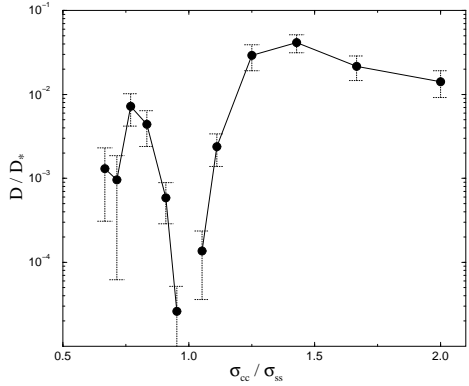}}
\vspace*{1cm}
\caption{Dependence of the diffusion coefficient on the mismatch between the lattice parameter of the substrate and
 the cluster. A small change
in the lattice parameter of the cluster lead to a huge change 
in the diffusivity.(N=100, $\epsilon_{sc}$=0.4, T=0.3, Run Length = 12500 $\tau$)
\label{}
}
\end{figure}

\begin{figure}
\vspace*{2cm}
\epsfxsize=6cm
\rotate[r]{\epsfbox{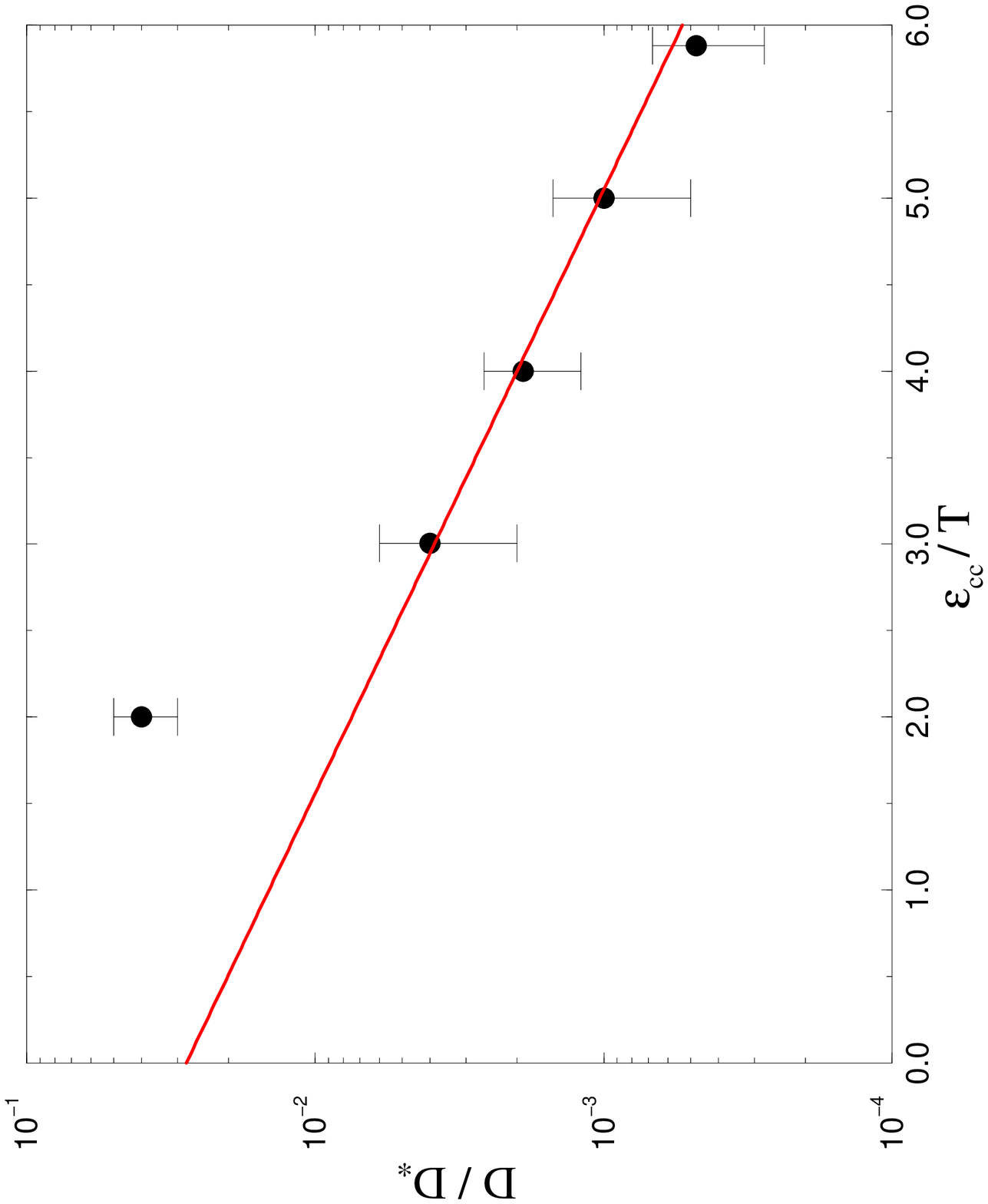}}
\vspace*{1cm}
\caption{Temperature dependence of the diffusion coefficient. The diffusion
follows an Arrhenius law.
\label{}
}
\end{figure}

\begin{figure}
\vspace*{2cm}
\epsfxsize=6cm
\rotate[r]{\epsfbox{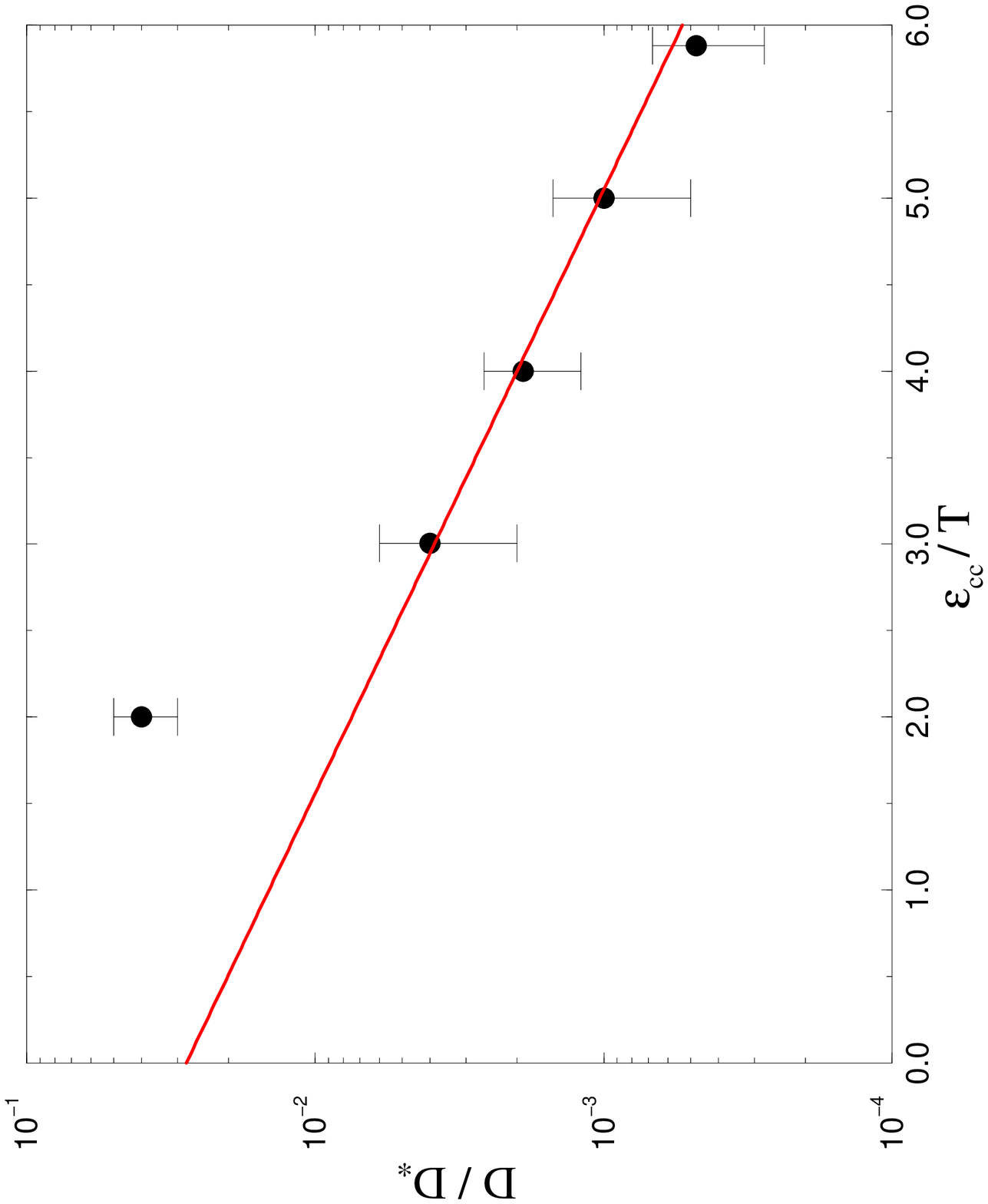}}
\vspace*{1cm}
\caption{Dependence of the diffusion coefficient of a cluster as a
function of its number of particles.
Data correspond to different
mismatches between the cluster and the substrate lattice parameters.
The diffusion coefficient decreases as a power law with exponent $\alpha$.
The two different slopes correspond to different diffusion regimes :
 the weaker dependence corresponds to a brownian trajectory; the stronger 
correspond to a ``hopping-like'' diffusion.
For comparison, the arrow indicates the diffusion coefficient of
a single adatom with $\sigma_{ss}=0.9$.
($\epsilon_{sc}$=0.4, N=100, Run Length = 12500 $\tau$)
\label{}
}
\end{figure}

\end{document}